\documentclass[conference, letterpaper]{IEEEtran}
\IEEEoverridecommandlockouts
\usepackage{cite}
\usepackage{amsmath,amssymb,amsfonts}
\usepackage{algorithmic}
\usepackage{graphicx}
\usepackage{textcomp}
\usepackage{xcolor}
\usepackage{todonotes}
\usepackage{flushend}
\usepackage[inline]{enumitem}
\def\BibTeX{{\rm B\kern-.05em{\sc i\kern-.025em b}\kern-.08em
    T\kern-.1667em\lower.7ex\hbox{E}\kern-.125emX}}
\begin{document}

\title{Trustworthy AI in the Age of Pervasive Computing and Big Data}

\author{\IEEEauthorblockN{Abhishek Kumar\IEEEauthorrefmark{1},
Tristan Braud\IEEEauthorrefmark{2}, Sasu Tarkoma\IEEEauthorrefmark{1},
Pan Hui\IEEEauthorrefmark{1}\IEEEauthorrefmark{2} }
\IEEEauthorblockA{\IEEEauthorrefmark{1}Department of Computer Science, University of Helsinki, Finland\\
\IEEEauthorrefmark{2}Department of Computer Science and Engineering, Hong Kong University of Science and Technology, Hong Kong\\
Email: abhishek.kumar@helsinki.fi, braudt@ust.hk, sasu.tarkoma@helsinki.fi, pan.hui@helsinki.fi}}

\maketitle

\begin{abstract}
The era of pervasive computing has resulted in countless devices that continuously monitor users and their environment, generating an abundance of user behavioural data. Such data may support improving the quality of service, but may also lead to adverse usages such as surveillance and advertisement. 
In parallel, Artificial Intelligence (AI) systems are being applied to sensitive fields such as healthcare, justice, or human resources, raising multiple concerns on the trustworthiness of such systems. 
Trust in AI systems is thus intrinsically linked to ethics, including the ethics of algorithms, the ethics of data, or the ethics of practice. 
In this paper, we formalise the requirements of trustworthy AI systems through an ethics perspective. We specifically focus on the aspects that can be integrated into the design and development of AI systems. 
After discussing the state of research and the remaining challenges, we show how a concrete use-case in smart cities can benefit from these methods. 
\end{abstract}


\section{Introduction}

Recent technical advances in computing and communication have led to a multiplication of devices embedding computational capabilities, a phenomenon more commonly known as pervasive computing. Such devices constantly monitor and sense users and their environment, producing a vast amount of behavioural data~\cite{haddadi2011targeted}. Emerging technologies such as Augmented Reality heavily rely on continuous video feeds of the users' surroundings~\cite{7979812}. Artificial Intelligence (AI) systems can exploit this data to learn more about users. This data collection and interpretation may be directed towards improving the quality of service, but may also serve other purposes, including surveillance, advertisement, and the algorithmisation of behaviours.
AI has started to reach
domains with a direct impact on human life, including justice, healthcare, and autonomous driving. In such fields, every decision can have dramatic outcomes, and there is no room for erroneous  conclusions.
However, AI systems are subject to various 
distortions which may lead to unfair decisions.

As AI systems vastly depend on data, AI-assisted decisions are only as right as the data provided for training. Such data often reflects existing biases in gender, race, or religion. The resulting AI systems will thus reinforce existing discriminations. Recent examples include recruitment algorithms that, being provided data containing a majority of male applicants, penalise resumes from female applicants~\cite{amazon2018}.
Besides the initial data provided for training, some AI systems rely on a continuous data stream for learning.
Such systems may be hijacked by malicious users to alter their original purpose. A famous real-world example is Tay, Microsoft's AI-powered chatbot, which started posting offensive tweets after interacting with Twitter users~\cite{tay2016}. In addition to existing bias in data, the usage of AI and data themselves may be questionable, as shown by the scandal of Cambridge Analytica influencing voters based on their Facebook profile~\cite{cambidgeanalytica2018}. Finally, some of the techniques behind AI such as deep neural networks, boosting, and random forest function as ``black boxes", making decisions without exposing the underlying reasons, and preventing their application in the most sensitive domains. It is hard to explain how these models behave, how they produce predictions, and what the influencing variables are. At the other end of the spectrum, white-box models such as linear regression and decision tree solve most of the above questions, at the cost of lower accuracy than black-box models. 
Such examples raise the question of public trust in often unregulated AI systems.

\begin{figure}[t]
    \centering
    \includegraphics[width=.4\textwidth]{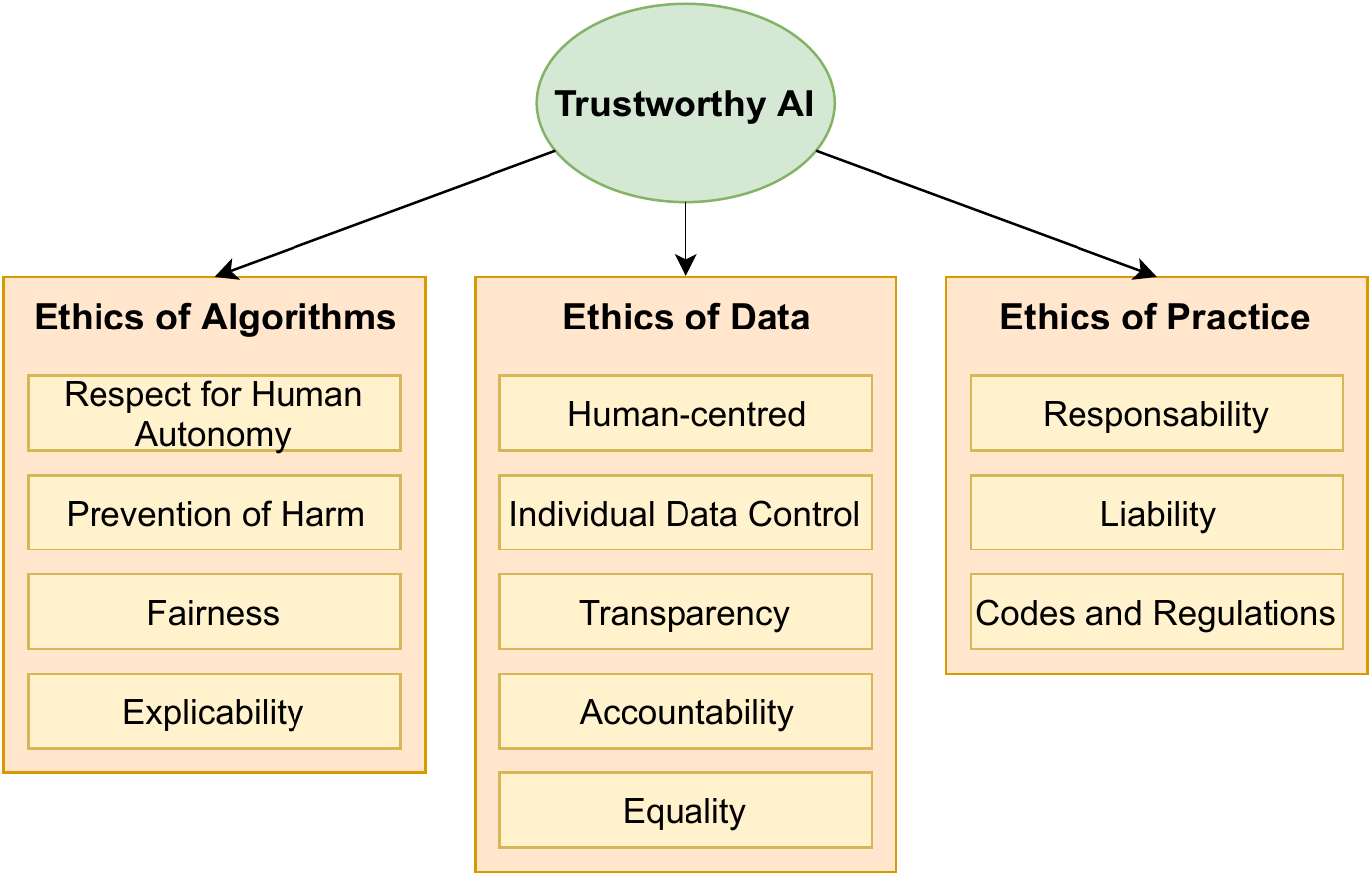}
    \caption{The three main components of a trustworthy AI~\cite{floridi2016data}.}
    \label{fig:trust}
    \vspace{-1em}
\end{figure}

Trustworthy AI can be seen as the sum of three components, shown in Figure~\ref{fig:trust}: ethics of algorithms, ethics of data, and ethics of practice~\cite{floridi2016data}. These  components provide a data-centric level of abstractions for ethical questions. Attempting to solve ethical issues for AI systems raises many open issues. What constitutes human ethics 
can hardly be summarised by typical representations such as decision trees. Transferring human ethics 
into a clearly defined machine ethics ruleset is thus an intricate problem. In this paper, we target the issues that can be solved at the development stage to provide trustworthy-by-design AI systems. As such, we focus primarily on the ethics of algorithms and ethics of data.
The current pervasiveness of computing resources and recent advances in federated learning allow designing distributed systems addressing many of the current challenges in AI ethics. After discussing the requirements of a trustworthy AI, we extensively review the current state of research on such issues and the remaining challenges. 
We finally describe how a concrete use-case in smart cities can benefit from such techniques.

\begin{figure*}
    \centering
    \includegraphics[width=.65\textwidth]{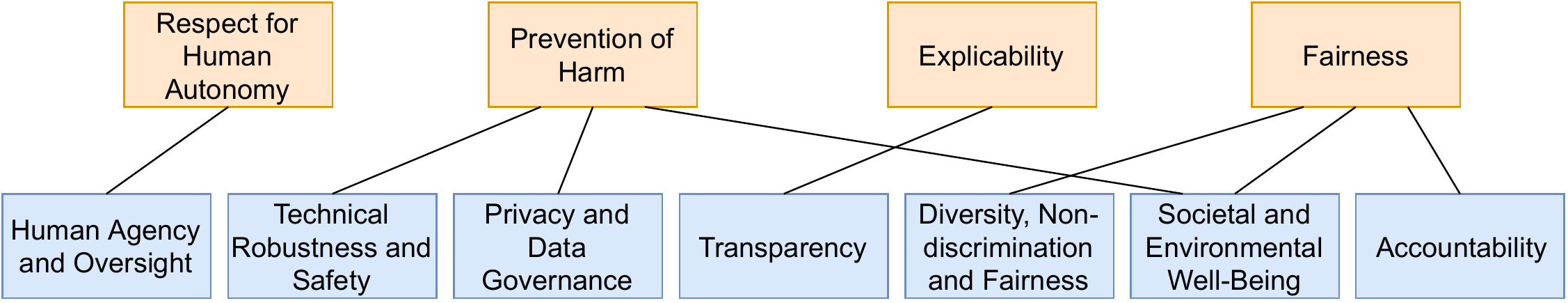}
    \caption{Ethics of Algorithms~\cite{hleg2019ethics}.}
    \label{fig:ethicsalgo}
    \vspace{-1em}
\end{figure*}

\section{Ethics of Algorithms}
In 2018, the High-Level Expert Group on AI set up by the European Commission released ethical guidelines for building trust in human-centric AI, towards Trustworthy AI~\cite{hleg2019ethics}. Such an AI system should respect the  following ethical principles: 
\begin{enumerate*}
    \item Respect for Human Autonomy
    \item Prevention of Harm
    \item Fairness
    \item Explicability.
\end{enumerate*}
Based on these four principles, the guidelines propose seven core requirements (Figure~\ref{fig:ethicsalgo}).

\subsection{Requirements for Trustworthy AI}\label{trustworthyai}

\subsubsection{Human agency and oversight}
AI systems should 
enforce the \emph{principle of respect for human autonomy}. AI systems should act as enablers to a democratic and equitable society by supporting the user's agency, foster fundamental rights, and allow for human oversight.

\subsubsection{Technical Robustness and Safety}
Technical robustness is related to \emph{prevention of harm}. It requires AI systems to be developed with a preventive approach to risks so that they  behave reliably while minimising harm. 
This should also apply to potential changes in their operating environment or the presence of adversaries attacking the system. 

\subsubsection{Privacy and Data Governance} 
Privacy is also closely linked to \emph{prevention of harm}, as a fundamental right particularly affected by the pervasive data collection behind AI systems. Prevention of harm to privacy also necessitates data governance that covers the quality and integrity of the data used, its relevance,
its access protocols and the capability to process data in a manner that protects privacy.

\subsubsection{Transparency}
This requirement is closely linked to \emph{explicability}. It seeks the transparency of all elements relevant to an AI system: the data, the system and the business models. In the age of pervasive computing, transparency is critical to justify extensive data collection and its benefits to users.


\subsubsection{Diversity, Non-discrimination and Fairness}
In order to achieve Trustworthy AI, we must enable inclusion and diversity throughout the entire AI system. Besides the consideration and involvement of all affected stakeholders, this also entails ensuring equal access 
and equal treatment. This requirement is closely related to \emph{fairness}.

\subsubsection{Societal and Environmental Well-being}
In line with \emph{fairness and prevention of harm}, the broader society and the environment should also be considered as stakeholders. Sustainability and ecological responsibility of AI systems should be encouraged, and research should be fostered into AI solutions addressing areas of global concerns.  AI systems should benefit all human beings, including future generations.


\subsubsection{Accountability}
The requirement of accountability is closely linked to the \emph{principle of fairness}. It necessitates mechanisms to ensure responsibility and accountability for AI systems and their outcomes, both before and after their development, deployment and use.

\subsection{Recent Research Towards Trustworthy AI System}

Many research works are tackling the seven requirements mentioned in Section~\ref{trustworthyai} for a trustworthy AI. In this section, we discuss the current status of research and the remaining challenges for each of the seven requirements.






\subsubsection{Human Agency and Oversight}~\\
\textbf{Human-in-the-Loop AI: } Originally, the Human-in-the-Loop AI approach was proposed as a workflow where AI learns from the human operator while intuitively making the human's work more efficient. Active learning is one such approach where humans provide labels for some unlabelled data to order to achieve the desired accuracy quickly. It can be utilised in designing AI systems. Recently, this approach has been exploited to design fairer paradigms for AI systems, i.e. AI systems which generate revenues will repay the legitimate owners of the knowledge used for taking those decisions~\cite{zanzotto2019human}.

\noindent\textbf{Meaningful Human Control (MHC): }In Autonomous Weapon Systems, the MHC paradigm ensures that humans have the power to influence or direct the course of events as well as the ability to manage a machine \cite{verdiesen2018design}. Using this paradigm to design AI systems can ensure the capability for human intervention during both the design and the operation, including the possibility to override a decision made by the AI system when it violates the law.

\subsubsection{Technical Robustness and Safety}~\\
\textbf{Adversarial Artificial Intelligence: }AI models are susceptible to various kinds of attacks, i.e. poisoning attacks~\cite{biggio2011support}\cite{floridi2016data}, evasion attacks~\cite{goodfellow2014explaining}, and model stealing attacks~\cite{tramer2016stealing}. Current state-of-the-art defence methods against these attacks focus on increasing the robustness of the model by injecting adversarial examples into the training set~\cite{DBLP:journals/corr/GoodfellowSS14}, hiding the model's information from adversaries~\cite{DBLP:conf/iclr/TramerKPGBM18}, reducing the sensitivity of the model by reducing its complexity~\cite{DBLP:conf/ndss/Xu0Q18} or minimising the transferability (especially in neural networks)~\cite{hosseini2017blocking}.

\noindent\textbf{Safe and Reliable Artificial Intelligence: }
 Reliability in AI systems is ensured by borrowing three principles: 1) Failure Prevention, 2) Failure Identification \& Reliability Monitoring, and 3) Maintenance. To prevent failures in AI systems, the current state-of-the-art methods try to proactively identify likely sources of error resulting from 1) bad or inadequate data~\cite{buolamwini2018gender}, 2) differences or shifts in environment~\cite{DBLP:conf/aistats/SubbaswamySS19}, 3) model associated errors~\cite{goodfellow2014explaining}, or 4) poor reporting~\cite{DBLP:conf/fat/MitchellWZBVHSR19} and develop methods that correct for these in advance. After deployment, reliability mechanisms assess the model output for each new input and reject the unreliable output based on the auditing criteria of the density principle and the local fit principle~\cite{schulam2019can}. Model maintenance requires detecting when updates to the model are necessary; however, unlike in traditional software engineering systems, the maintenance cost is already compounded due to development complexity~\cite{43146}.


\subsubsection{Privacy and Data Governance}~\\
\textbf{Privacy by Design: } The privacy by design approach calls for privacy concerns to be predominant throughout the whole engineering process. It was originally designed for traditional software systems. In the context of AI systems, it entails seeking explicit consent from data owners before using their data in training the model and respecting clauses like the ``right to be forgotten" under General Data Protection Regulation (GDPR), a regulation in EU law on data protection and privacy in the European Union and the European Economic Area, whenever invoked by the data contributors. Efficient solution for such implementing such clauses in AI systems has just begun to be proposed and largely remains unexplored~\cite{cao2015towards}\cite{barua2016time}.


\noindent\textbf{Federated Architecture: }Federated architectures enable training AI systems without collecting users' personal data at a centralised location. With federated learning~\cite{mcmahan2017communication} or peer-to-peer (P2P) machine learning\cite{bellet2018personalized}, users do not have to share raw data, rather only training updates. Differential privacy~\cite{dwork2014algorithmic} and secure aggregation~\cite{bonawitz2017practical} enforce another layer of privacy to make training updates more secure.


\subsubsection{Transparency}~\\
\textbf{Explainable Artificial Intelligence: }Explainability in AI systems requires that the decisions made by an AI system can be understood and traced by human beings. Explainability in traditional AI systems (using rule/tree-based models like decision tree, linear models like linear regression, Gaussian mixture model) is straightforward because the decision boundaries corresponding to these models are easy to visualise. Explainability in deep learning  AI systems is not possible due to its non-linear structure (hence, known as black-box models). Current attempts of explainability for black-box models are focused on input attribution~\cite{DBLP:journals/corr/abs-1906-02825}, concept testing/extraction~\cite{ghorbani2019towards}, example influence/matching~\cite{DBLP:journals/corr/abs-1902-06292}, and distillation~\cite{hinton2015distilling}.

\noindent\textbf{Communication: }AI-based Chatbots like Google Duplex can mimic human sound so perfectly that even a human could not tell whether they were talking to a robot (may be used for telephone-based scams\cite{li2018machine}). Generative adversarial networks can produce fake images which look real to human users~\cite{radford2015unsupervised}\cite{karras2019style}. Transparency by communication stresses that AI systems should not present themselves as humans to users and humans have the right to be informed that they are interacting with an AI system. 
Efficient solutions to distinguish human initiated action from machine initiated action remain unexplored.   

\subsubsection{Diversity, Non-discrimination and Fairness}~\\
\textbf{Discrimination Discovery: }Discovery aims at finding discriminatory patterns in data using data mining and machine learning methods~\cite{ruggieri2010data}\cite{mancuhan2014combating}. It builds upon extensive research in statistics on discrimination evidence~\cite{tinkham2010uses}, addressing new challenges due to the increasing volumes and complexity of data and ways of possible unfairness. Statistics has been focusing on hypotheses testing in decision data and provide essential solutions to compare groups of people correctly.

\noindent\textbf{Discrimination Prevention: }Discrimination prevention develops methods for sanitising algorithms or adjusting machine learning processes so that outputs obey the fairness constraints. Several attempts to fix algorithms include prepossessing training datasets ~\cite{feldman2015certifying}, adding a regularizer to the model~\cite{zemel2013learning}, post-processing trained models~\cite{6175897} or
model outputs~\cite{hajian2012methodology}.


\subsubsection{Societal and Environmental Well-being}~\\
\textbf{Computational Sustainability: }Computational sustainability requires the development and deployment of AI systems to tackle pressing societal concerns in the most environmentally friendly way possible. It involves efficient resource usage and energy consumption during training and deployment with minimum carbon footprint~\cite{fisher2017selected}\cite{khakurel2018rise}. In smart cities, AI-based transportation systems plan efficient travel routes while minimising greenhouse gas emissions for  commuters ~\cite{guerrero2015integration}.

\noindent\textbf{Affective Computing: }Ubiquitous exposure to social AI systems due to wide-scale usage of wearables and social media in many of our lives can alter our conception of social agency, or impact our social relationships and attachment. For example, algorithms managing the newsfeed of  Facebook users may influence the users' political perception~\cite{alvarado2018towards}. Affective computing advocates computers with empathy and giving emotional intelligence to machines~\cite{picard2000affective}. 

\subsubsection{Accountability}~\\
\textbf{Data Provenance: }Data provenance methods track the flow of data from end to end, across technical and administrative boundaries, thereby bringing accountability by providing evidence—for example, how personal data is collected and subsequently processed, improper behaviour such as unjustified personal data disclosure to an advertiser~\cite{singh2018decision}\cite{pasquier2018data}. 

\noindent\textbf{Data Auditing: }Data auditing for AI systems takes place during both development and deployment. During the development, software verification methods~\cite{wallace1989software} ensure compliance with regulations and company policies. For example, GDPR mandates privacy (and data protection) by design and by default, i.e. seeking explicit consent from users and applying the strictest privacy settings by default. During deployment, blockchain or distributed ledger approaches allow to establish ``immutable" records that can operate without requiring trusted third parties and hence maintaining integrity~\cite{dai2017toward}. 



\section{Ethics of Data and Ethics of Practice: Towards Trustworthy AI Systems}

AI systems differ from traditional decentralised systems in one significant aspect: AI models revolve around data. 
It is one of the main reasons why companies with an active AI engagement, such as Google, Amazon, or Facebook, collect users' personal information pervasively. This data collection includes information actively shared by the user, but also a multitude of other parameters sensed by other devices. 
In return, users benefit from their services for free. This arrangement can be a win-win scenario for both parties. However, some recent abuses threaten the implicit understanding of this arrangement. One may remind the supermarket Target figuring out the pregnancy of a teenager before her father~\cite{target2012}, 
or Cambridge Analytica influencing voters based on their Facebook profile~\cite{cambidgeanalytica2018}. Such scandals call for a set of measures on data ethics, that is the responsible and sustainable use of data by companies, authorities and organisation to sustain user's trust. DataEthics, a Denmark-based politically independent ThinkDoTank, recommends five principles: \textbf{1) Human being at the center, 2) Individual data control, 3) Transparency, 4) Accountability, and 5) Equality}~\cite{dataethics2018} to enforce data ethics.


\subsection{Ethics of Data}
The ethics of data focuses on ethical problems posed by the collection and analysis of large datasets using AI or Data Mining techniques which makes it possible to re-identify individuals via linking with other auxiliary datasets.  
\textcolor{black}{With the advent of 5G, sensing and AI are becoming more pervasive, supported by edge computing, in-network AI, augmented reality, and the Internet of Things~\cite{saghezchi2015drivers}. This phenomenon leads to the problem of \emph{ethics of sensing}. Where can we place the logic that guides and governs the ethical behaviour of a system or application? Should some of the logic be in the sensor itself to prevent misuse?}
Two directions can enforce data ethics. A first direction lies in using design principles which advocate storing very little or virtually no amount of user's raw data. The other direction is letting users decide which data they want to share and even earn economic benefits~\cite{laudon1996markets}.
This approach aims to give control of personal data back to users. It can play a significant role in challenging the current status-quo of \emph{Surveillance Capitalism}, i.e. the commodification of users' personal data and their transformation into behavioural data for analysis and sale. 

\subsubsection{Design Principles for Ethics of Data}

\begin{itemize}

    \item \textbf{Zero Knowledge as a Design Principle: }According to the GDPR, no data must be stored longer than is necessary. Companies can  go beyond the legislation and delete data before the required date. It can be done through auto-deletion, but also by never having access to the data in the first place. Federated learning or P2P based AI systems are an example of such design principles where the user's raw data never leaves their device~\cite{mcmahan2017communication}.
    
    \item \textbf{Contextual Integrity as a Design Principle: }Many organisations collect data on people's lives and activities without their knowledge~\cite{zimmer2017internet}. In 2016, a group of Danish researchers publicly released a dataset of nearly 70,000 users of the online dating site OkCupid, including usernames, age, gender, location, sexual orientation, personality traits, and answers to thousands of personal profiling questions~\cite{danish2016}. Although such information was already public on OkCupid, users did not intend for a study to exploit their data. The Facebook–Cambridge Analytica data scandal is another emblematic example. The theory of Contextual Integrity allows enforcing data ethics by providing a framework for evaluating the flow of personal information between agents to identify and explain why certain patterns of information flow are acceptable in one context but viewed as problematic in another~\cite{nissenbaum2004privacy}\cite{doi:10.1177/2056305118768300}.
\end{itemize}


\subsubsection{Personal Data Infrastructure for Ethics of Data}
\begin{itemize}
    \item \textbf{Personal Databox: } The idea of a personal databox was proposed by Chaudhry et. al in 2015~\cite{DBLP:conf/aarhus/ChaudhryCHMMHM15}. Personal databoxes act as a single-point (physical) gateway through which users' data flows to anywhere outside the user's control. It allows users to capture, index, store and manage data about them as well as data generated by themselves. 
    \item \textbf{Hub-of-All-Things (HAT): }HAT provides users with a personal data infrastructure. Unlike Databox, HAT relies on the idea that users should earn economic value from their data~\cite{ng2018market}. Users generate durable data and perishable data. 
    Perishable data is often not used within its expiry window and hence, loses its value. HAT allows capturing such data and brokers benefits from their parties available in its ecosystem. HAT advocates the fact that users should receive explicit benefits for their data (privacy trading).
\end{itemize}

\subsection{Ethics of Practice}
Ethics of practices focuses on the responsibilities and liabilities of people and organisations in charge of data. It aims to determine and maintain guarantees for the ethical use of data and models when disclosed at multiple stages and aggregated by multiple parties. To this purpose, it is necessary to define both a technical~\cite{zhang2018blockchain} and ethical framework to guide decisions such as data release without getting into scandals like the AOL search data leak~\cite{chiru2016search} or Netflix privacy lawsuit~\cite{netflix2010}. More recently, it has also extended to the practice of model and synthetic model disclosure, since models trained on user's data may release private information using new types of attacks like membership inference attack~\cite{7958568}. 

\section{Use-case: Trustworthy AI in Smart Cities}

Smart cities are a concrete example of pervasive computing and data sensing. A multitude of sensors collect data for AI models to provide insights on traffic management and road safety~\cite{8730706}, infrastructure monitoring, or community service planning. However, such data collection will inherently expose users' private data to a risk of misuse by authorities and companies, such as location profiling.

 
 Consider the following scenario:
 \textit{``A city council plans to install chargers for electric vehicles and needs spatiotemporal information from citizens' vehicles. The number and the location of the chargers will depend on the collected information. Citizens are invited to participate in the data collection process."} 
 The city council has two options: 
 \begin{enumerate*}
     \item collect and anonymise citizens' data
     \item build an AI system which will recommend the number and the locations of the installations.
 \end{enumerate*}

\begin{figure}[t]
    \centering
    \includegraphics[width=.35\textwidth]{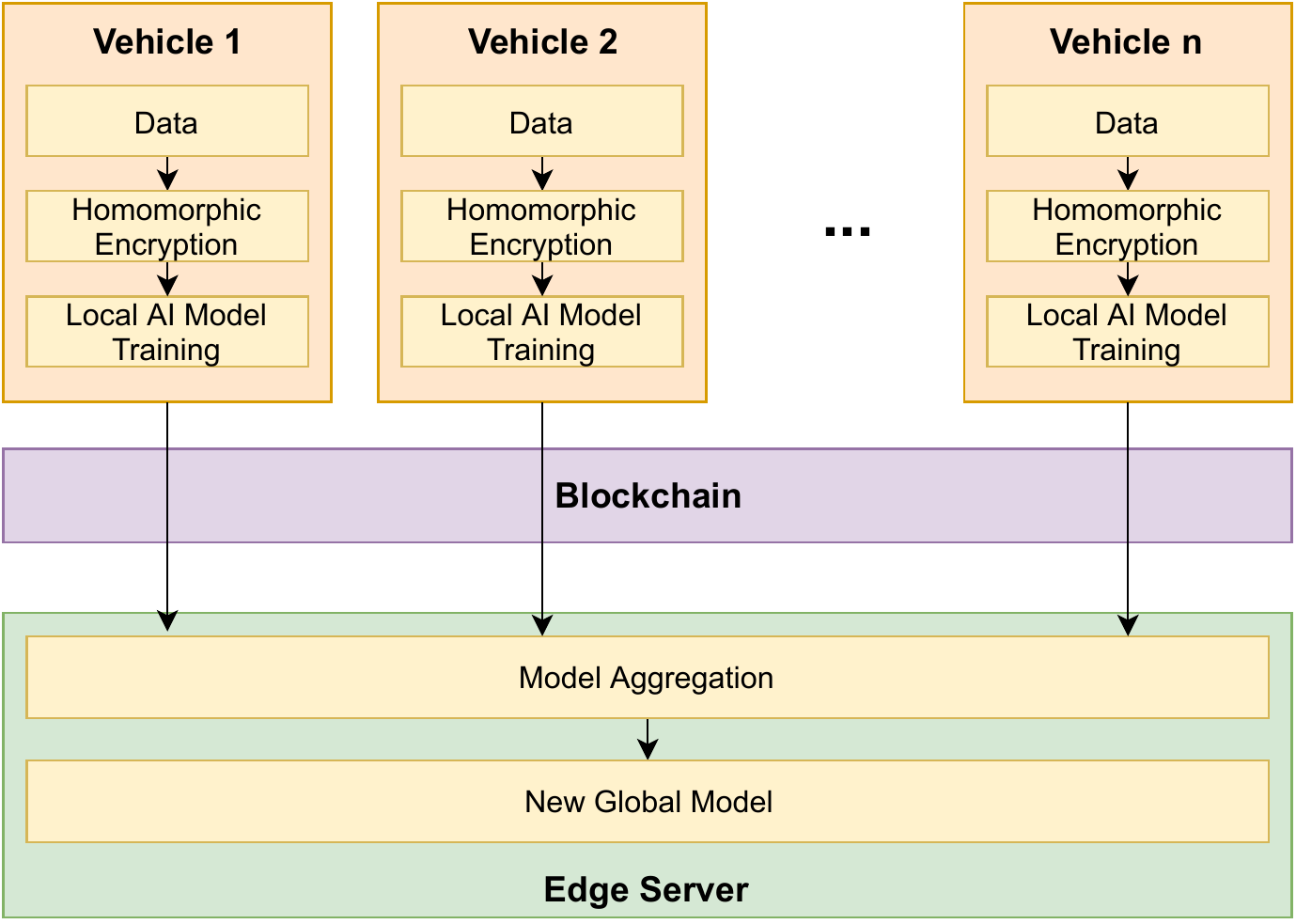}
    \caption{Trustworthy AI in a smart city scenario.}
    \label{fig:my_label}
    \vspace{-2em}
\end{figure}
Recent advances in data mining techniques have made deanonymisation obsolete. For example, De Montjoye et al. showed that four spatiotemporal points are enough to uniquely identify 95\% of people in a mobile phone database of 1.5M people~\cite{de2013unique}. Building an AI system based \textcolor{black}{on on-device learning, such as federated learning}, protects against these problems since the data never leaves the vehicle. Given that vehicles have enough computing power, using federated learning in smart cities' vehicular networks for predicting the number and the locations of the charger installations is technically feasible while protecting users' privacy. 
Since federated learning systems rely on a zero-knowledge design, they can address the \textit{ethics of data} by on-device processing, i.e. raw data of users do not leave their device. An additional layer of privacy can be enforced with differential privacy or homomorphic encryption in the user's training algorithm. Such a system combined with a distributed ledger such as blockchain allows for accountability as each model update gets logged as an immutable record \textcolor{black}{(as shown in Figure~\ref{fig:my_label})}. \textcolor{black}{With the increasing number of users and contributors, the consensus mechanism responsible for updating the global model will become computationally expensive, thus costing more energy and preventing real-time applications. A potential solution lies within adding an extra layer to the Blockchain to address the scalability issue. For instance,~\cite{8306880} adds a Management hub to the Blockchain network to tackle access control problem in the presence of billions of IoT devices.} Finally, the distributed nature of federated learning allows for resilience to system failure. On the other hand, the architecture alone cannot address all the requirements. Fairness, explainability, and resilience against evasion attacks can be achieved through AI algorithms that tackle the black-box nature of current state-of-the-art AI algorithms, bringing more transparency to the system. 



\section{Conclusion}
In this paper, we proposed a framework for trustworthy AI systems in the age of pervasive data collection and computing. This framework provides a data-centric level of abstractions for ethical questions posed in the AI and Data Science context. This Data-centric level of abstractions provides ethical abstractions on there level: data, algorithm, and practice. 
We focused on the ethics of data and ethics of algorithms, and more specifically, the aspects that can be integrated directly within the design and development of AI systems. After reviewing the challenges and requirement, as well as the current status of research, we discussed how a concrete use-case in the context of smart cities could benefit from the existing techniques for a more trustworthy usage of AI.

\section*{Acknowledgment} 
This   research   has   been   supported   in   part   by   project 16214817  from  the   Hong  Kong Research  Grants  Council, the 5GEAR project and the FIT project from the Academy of Finland.



\bibliographystyle{IEEEtran}
\bibliography{references}

\end{document}